\documentclass[aps,prb,superscriptaddress,amsmath,amssymb,preprintnumbers,showpacs]{revtex4}
\usepackage{graphicx}
\newcommand{\bra}[1]{\langle{#1}|}
\newcommand{\ket}[1]{|{#1}\rangle}
\newcommand{\braket}[2]{\langle{#1}|{#2}\rangle}

\begin{document}
\title{On the generation of multipartite entangled states in Josephson architectures}
\author{Rosanna Migliore}\email{rosanna@fisica.unipa.it}
\affiliation{CNR-INFM, MIUR, and Dipartimento di Scienze Fisiche
ed Astronomiche dell'Universit\`a di Palermo, Via Archirafi 36,
I-90123 Palermo, Italy}
\author{Kazuya Yuasa}\email{yuasa@hep.phys.waseda.ac.jp}
\altaffiliation[Present address: ]{Research Center for Information Security, National Institute of Advanced Industrial Science and Technology (AIST), 1-18-13 Sotokanda, Chiyoda-ku, Tokyo 101-0021, Japan}
\author{Hiromichi Nakazato}\email{hiromici@waseda.jp}
\affiliation{Department of Physics, Waseda University, Tokyo 169-8555, Japan}
\author{Antonino Messina}\email{messina@fisica.unipa.it}
\affiliation{MIUR and Dipartimento di Scienze Fisiche ed
Astronomiche dell'Universit\`a di Palermo, Via Archirafi 36,
I-90123 Palermo, Italy}
\date[]{April 11, 2006}

\begin{abstract}
We propose and analyze a scheme for the generation of multipartite
entangled states in a system of inductively coupled Josephson flux
qubits. The qubits have fixed eigenfrequencies during the whole
process in order to minimize decoherence effects and their
inductive coupling can be turned on and off at will by tuning an
external control flux. Within this framework, we will show that a
W state in a system of three or more qubits can be generated by
exploiting the sequential one by one coupling of the qubits with
one of them playing the role of an entanglement mediator.
\end{abstract}
\maketitle
\section{Introduction}
\label{intro} Entanglement ``the striking feature of quantum
mechanics'' (as claimed by E. Schr\"odinger in 1935 \cite{schro})
has been considered essential since the very beginning in order to
investigate fundamental aspects of quantum theory. Quite recently,
however, physicists have fully recognized that the generation of
entangled states is an essential resource also in quantum
communication and information processing. Entangled states have
been generated in many experiments involving cavity QED and NMR
systems, ion traps, and solid-state (superconducting) circuits,
and their applications in the field of quantum computing have been
demonstrated.\cite{cavity,multiqubit,dot,ions,zanardi} Among the
previously mentioned physical systems, Josephson-junction based
devices presently provide one of the best qubit candidates for the
realization of a quantum computer due to the fact that a wide
variety of potential designs for qubits and their couplings are
available and that qubits can be easily scaled to large arrays and
integrated in electronic circuits.\cite{makhlin,noritoday} A
series of successfully performed experiments for charge, flux,
phase, and charge-flux qubits show indeed that they satisfy
DiVincenzo's prescriptions \cite{divincenzo} for quantum computing
in terms of state preparation, state manipulation, and readout.
Moreover, the nonlinearity characterizing Josephson junctions and
the flexibility in circuit layout offer many possible options for
coupling qubits together and for calibrating and adjusting the
qubit parameters over a wide range of values.

In this field, remarkable achievements include the realization
of complex single-qubit manipulation schemes,\cite{[3]} the generation
of entangled states \cite{[4],[5]} in systems of coupled flux \cite{[3a]}
and phase \cite{[4a],[5a]} qubits, as well as the observation of quantum coherent
 oscillations and conditional gate operations using two
 coupled superconducting charge qubits.\cite{[2],[4]}
The next major step toward building a Josephson-junction based quantum computer
 is therefore to experimentally realize simple quantum algorithms, such as the
  creation of an entangled state involving more than two coupled qubits.\cite{zanardi,multiqubit,w,cirac,yuasa}

This goal may be achieved by selectively turning on and off the
direct couplings between two qubits or their interaction with
auxiliary systems ($LC$-oscillator modes,\cite{buisson}
inductances,\cite{nori} large-area current-biased Josephson
junctions,\cite{martinis,wei} or the quantized modes of a resonant
microwave cavity \cite{noi-prb,girvin,frunzio}) playing the role
of a data bus. Typically, the coupling energy may be controlled by
tuning the qubit level spacings in and out of resonance. However,
in order to avoid introducing extra decoherence with respect to
that characterizing single-qubit operations, other promising
scheme for realizing a tunable coupling of superconducting
(spatially separated) qubits have emerged: for instance that
wherein the interaction between two flux-based qubits is
controlled by means of a superconducting transformer with variable
flux transfer function,\cite{cosmellitransf} or those wherein two
qubits with an initial detuning can be made to (resonantly)
interact by applying a time-dependent (microwave) magnetic flux to
the qubits.\cite{norideph}

Within these experimental frameworks, we propose a theoretical
scheme by which it is possible to entangle \textit{more than two}
(spatially separated) flux qubits. It is based on the sequential
inductive interaction of the qubits with one of them acting as an
entanglement mediator.\cite{mediator,Haroche} More in detail, we
will see that the scheme operates in such a way that it generates
an entangled W state after a finite number of steps, that no
conditional measurement is required, and that the proposed
architectures are scalable, at least in principle, to an arbitrary
number of qubits.

The paper is organized as follows. First, in Sec.\
\ref{sec:FluxQubits}, we briefly describe the main features and
the Hamiltonians characterizing two kinds of Josephson devices,
namely the double rf-SQUID \cite{luk,cosm,chiarello} and the
persistent (three-junction) SQUID,\cite{mooij99}  by which it is
possible to build superconducting flux qubits. Then, we discuss
the most common experimental procedures by which it is possible to
initialize and to measure their quantum state. In Sec.\
\ref{sec:Scheme}, a scheme of successive interaction is introduced
in an ($N+1$)-qubit system, qubit M+qubit 1+$\cdots$+qubit $N$,
wherein entanglement mediator M is coupled one by one with qubits
1, 2, \ldots, $N$. We discuss moreover the possibility of
practically realizing this scheme by exploiting some of the
different physical coupling elements currently available and how
the coupling energy depends on the particular way in which the
interaction between each qubit and the mediator is implemented. In
Sec.\ \ref{sec:Analysis}, we analyze the dynamics of the system
showing that, by preparing the multi-qubit system in a pure
factorized state and by adjusting the coupling energies and/or the
time of interaction between each of them and the mediator, an
entangled W state can be generated. Finally, conclusions and
discussions are given in Sec.\ \ref{sec:Conclusion}.

\section{Superconducting Flux Qubits: Models and Hamiltonians}
\label{sec:FluxQubits} In this section, we briefly describe the
main features and the Hamiltonians of two devices, the tunable
rf-SQUID \cite{luk,cosm,chiarello} and the three-junction
SQUID,\cite{mooij99} by which it is possible to implement a
two-state Josephson system, focusing our attention also on the
experimental procedures to be considered in order to prepare their
initial quantum  state.

\subsection{The double rf-SQUID qubit}
\begin{figure}
\includegraphics[width=0.7\textwidth]{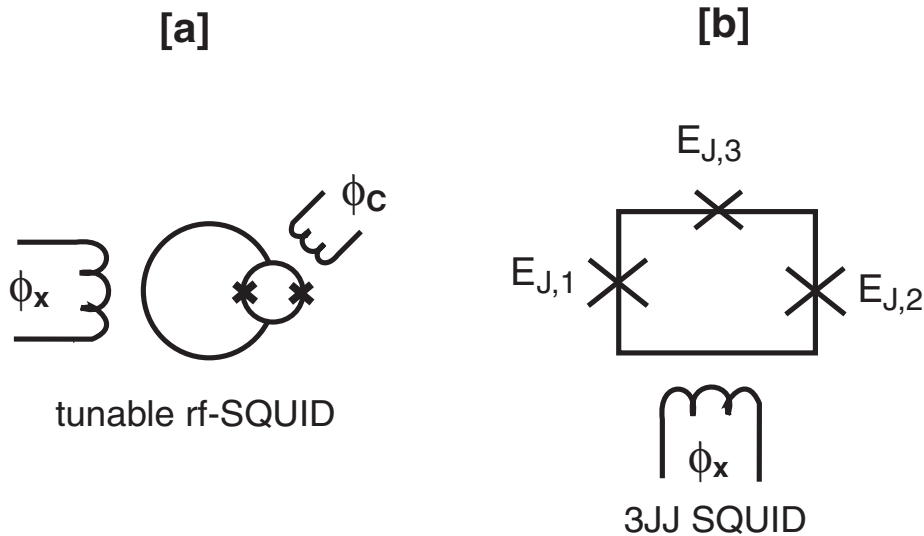}
\caption{Sketch (a) of the superconducting double rf-SQUID qubit
and (b) of the three-junction SQUID by which it is possible to
realize a persistent-current qubit.}\label{squids}
\end{figure}

We begin by considering a double rf-SQUID
system,\cite{luk,cosm,chiarello} that is a superconducting ring of
self-inductance $L$ interrupted by a dc-SQUID, a smallest loop
containing two identical Josephson junctions, each with critical
current $i_0$ and capacitance $c$. This device [schematically
illustrated in Fig.\ \ref{squids}(a)] is biased by two magnetic
fluxes $\phi_x$  and $\phi_c$ threading the greatest ring and the
dc-SQUID, respectively.

The dc-SQUID if small enough (i.e.\ with inductance $l\ll
i_0\phi_0/2\pi$, where $\phi_0=h/2e$ is the flux quantum) behaves
like a single junction with a flux-dependent critical current
$I_c=2i_0|\cos (\pi\phi_c/\phi_0)|$ and capacitance $C=2c$. This
means that a double SQUID simulates a standard rf-SQUID with
tunable critical current $I_c\equiv I_c(\phi_c)$.

Therefore, by taking into account both the charging energy of the
``effective dc-SQUID junction'' ($T=q^2/2C$) and the washboard
potential, the Hamiltonian of the system is written down as
\begin{equation}
H=\frac{q^2}{2C}
+\frac{(\phi-\phi_x)^2}{2L}
-\frac{I_c\phi_0}{2\pi}\cos\!\left(\frac{2\pi\phi}{\phi_0}\right),
\end{equation}
where the charge on the junction capacitance, $q$, and the flux
through the SQUID loop, $\phi$, are canonically conjugate
operators satisfying the commutation relation $[\phi,q]=i\hbar$.

It is well known \cite{chiarello,chiarello2} that, by setting
$\beta_L \equiv 2 \pi LI_c/\phi_0>1$ and $\phi_x\approx\phi_0/2$,
the circuit behaves as an artificial quantum two-level atom whose
reduced Hamiltonian in the basis of the flux eigenstates $\ket{L}$
and $\ket{R}$ (which are localized in the two minima of the
washboard potential and correspond to two different orientations
of the current circulating in the large loop) assumes the form
\begin{equation}\label{hsqui}
H_\text{RF}=-\frac{\hbar}{2}\Delta_\text{RF}(\phi_c)\sigma_x
-\frac{\hbar}{2}\epsilon_\text{RF}(\phi_x)\sigma_z.
\end{equation}
Here,
$\hbar\epsilon_\text{RF}(\phi_x)=2I_c\sqrt{6(\beta_L-1)}(\phi_x-\phi_0/2)$
is the energy difference between the two minima of the washboard
potential,
\begin{equation}\label{deltafi}
\Delta_\text{RF}(\phi_c)\simeq\frac{3}{2}\frac{\phi_0^2}{(2 \pi)^2L}
\left(1-\frac{\phi_0}{2\pi LI_c(\phi_c)}\right)^2
\end{equation}
(in the limit $0<\beta_L-1\ll1$) the tunnelling frequency between
the left and the right wells that can be tuned by changing the
junction critical current $I_c(\phi_c)$, and $\sigma_x$ and
$\sigma_z$ the Pauli spin operators.

\subsection{The persistent-current (3JJ) qubit}
To minimize the susceptibility to external noise of a
large-inductance rf-SQUID, Mooij \textit{et~al.}\cite{mooij99}
proposed to use a persistent-current qubit [schematically shown in
Fig.\ \ref{squids}(b)], namely a smaller superconducting loop
containing three Josephson junctions, two of equal size (i.e.\
with $E_{J,1}=E_{J,2}=E_J$) and the third smaller by a factor
$\alpha$ (i.e.\ with $E_{J,3}=\alpha E_J$, $\alpha<1$). By
applying an external magnetic flux $\phi_x$ close to $\phi_0/2$
and choosing $\alpha\approx0.8$, it has been proved that, in the
low-inductance limit (in which the total flux coincides with the
external flux and fluxoid quantization around the loop imposes the
constraint $\varphi_1-\varphi_2+\varphi_3+2\pi f=0$ on the phase
drops across the three junctions, $f=\phi_x/\phi_0$ being the
reduced magnet flux), the Josephson energy
\begin{equation}
U(\varphi_1,\varphi_2)=-E_J\cos\varphi_1-E_J\cos\varphi_2
-\alpha E_J\cos(2\pi f+\varphi_1-\varphi_2)
\end{equation}
forms a double well which permits two stable configurations of
minimum energy corresponding to two persistent currents of
opposite sign in the loop. This means that, also in this case, we
may engineer a two-state quantum system whose effective
Hamiltonian, in the basis of the two states that carry an average
persistent current $\pm I_p\approx\pm2\pi\alpha E_J/\phi_0$ (named
$\ket{L}$ and $\ket{R}$ also in this case), reads
\begin{equation}
H_\text{3JJ}=-\frac{\hbar}{2}\Delta_\text{3JJ}\sigma_x
-\frac{\hbar}{2}\epsilon_\text{3JJ}(\phi_x)\sigma_z.
\end{equation}
Here, the tunnelling matrix element between the two basis states,
$\hbar\Delta_\text{3JJ}/2$, depends on the system parameters, and
$\epsilon_\text{3JJ}(\phi_x)=2I_p(\phi_x-\phi_0/2)/\hbar$.

\subsection{Initialization and readout}
Using as a new basis that is spanned by the energy eigenstates
$\ket{0}$ and $\ket{1}$ (which are the symmetric and the
antisymmetric linear superpositions of $\ket{L}$ and $\ket{R}$, if
$\phi_x$ is \textit{exactly} equal to $\phi_0/2$), the
Hamiltonians of both the rf-SQUID qubit and the 3JJ qubit take the
diagonal spin $1/2$-like form
\begin{equation}
H=\frac{\hbar}{2}\omega \sigma_z.
\end{equation}
Here, $\hbar\omega\equiv E_1-E_0=\hbar\sqrt{\epsilon^2+\Delta^2}$
indicates the energy separation between their corresponding
eigenstates (the analytic form of $\epsilon$ and $\Delta$, as
previously discussed, depends on the specific design of the qubit)
and $\sigma_z=\ket{1}\bra{1}-\ket{0}\bra{0}$ is a Pauli operator.

Before discussing the \textit{modus operandi} of our scheme, it is
worth noting that both the rf-SQUID qubit and the 3JJ qubit are
easily addressed and measured. Usually, they are initialized to
the ground state simply by allowing them to relax, so that the
thermal population of their excited states can be neglected. The
coherent control of the qubit state is instead achieved via
NMR-like manipulation techniques, i.e.\ by applying resonant
microwave pulses which, by opportunely choosing the interaction
time and the microwave amplitude, can induce a transition between
the two qubit energy levels.\cite{[3]}

In addition, a flux state can be prepared producing a collapse of
the wave function through a flux measurement or, as recently
pointed out by Chiarello,\cite{chiarello2} in the case of a double
rf-SQUID qubit with an opportunely chosen sequence of variations
of the washboard potential.
 A double SQUID can be indeed prepared in  a particular flux state
by strongly unbalancing the washboard potential in order to have
just one minimum, then waiting a time sufficient for the
relaxation to this minimum and finally sufficiently raising the
barrier in order to freeze the qubit in this state. Finally,
coherent rotation between the two flux states can be realized by
lowering the barrier in order to induce fast free oscillations,
waiting for fractions of the oscillation period to realize the
desired rotation and opportunely raising the barrier in such a way
to freeze the system in the desired target state.

Also for qubit readout, several detectors have been experimentally
investigated. Most of them include a dc-SQUID magnetometer
inductively coupled to the qubit to be measured by which it is
possible to detect the magnetization signal produced by the
persistent currents flowing through it, exploiting the fact that
the dc-SQUID critical current is a periodic function of the
magnetic flux threading its loop.\cite{cosm} In addition, it is
worth emphasizing that, besides these proposals,  physicists have
been focusing their efforts on the realization of
nondemolition-measurement schemes (necessary for applications where
low back-action is required) like that based on the dispersive
measurement of the qubit state by coupling the qubit nonresonantly
to a transmission-line resonator and probing the transmission
spectrum.\cite{readout}

\section{The Scheme for Entanglement Generation: Sequential Interaction of $N$ Flux Qubits with an Entanglement Mediator}
\label{sec:Scheme} In this section, we propose a scheme for the
generation of maximally entangled states in a multi-qubit system,
$\text{M}+1+2+\cdots+N$, where qubit M, playing the role of an
entanglement mediator, is assumed to interact one by one with $1$,
$2$, \ldots, and $N$. Among the different forms of coupling
theoretically proposed and experimentally realized, we consider
those by which it is possible to realize an inductive interaction
\cite{cosmellitransf,norideph,plourde}  between each qubit and the
mediator, so that the free Hamiltonian of the whole system and the
interaction Hamiltonians can be cast (in the basis of the energy
eigenstates) in the following form:
\begin{equation}
H_0=\sum_{i=M,1,2,\ldots,N}\frac{\hbar}{2}\omega_i\sigma_z^{(i)}
\end{equation}
and
\begin{equation}
H_{M1}'=g_1\sigma_x^{(M)}\sigma_x^{(1)},\quad
H_{M2}'=g_2\sigma_x^{(M)}\sigma_x^{(2)},\quad\ldots,\quad
H_{MN}'=g_N\sigma_x^{(M)}\sigma_x^{(N)}.
\end{equation}
We assume moreover that these qubits are properly initialized, by
exploiting one of the previously mentioned experimental recipes,
so that at $t=0$ the whole system is described by the pure
factorized state
$\ket{1_M0_10_2\ldots0_N}\equiv\ket{1}_M\otimes\ket{0}_1\otimes\ket{0}_2
\otimes\cdots\otimes\ket{0}_N$.

With this setup, if qubits  $1$, $2$, \ldots, and $N$ are
spatially separated (in order to strongly reduce their direct
persistent coupling) and their interaction with mediator M can be
turned on and off at will (by adjusting the coupling energies
$g_1$, $g_2$, \ldots, $g_N$), we realize a step by step scheme
which is sketched as follows:
\begin{itemize}
\item  Mediator M prepared in the state $\ket{1}_M$ interacts
inductively (by setting $g_1\neq 0$) with qubit $1$ during an
opportunely chosen interval of time $0<t<\tau_1$, while qubits
$2$, $3$, \ldots, and $N$ evolve freely (with
$g_2=g_3=\cdots=g_N=0$). \item At $t=\tau_1$, we turn off the
interaction between M and $1$ and we adjust $g_2$ in order to
couple the mediator and qubit $2$ (by choosing $g_2\neq0$) for
$\tau_1<t<\tau_1+\tau_2$. \item In a similar manner, we put qubits
$3$, $4$, \ldots, and $N$ in interaction with M one by one. \item
Finally, at $t=\tau_1+\tau_2+\cdots+\tau_N$, we switch off the
interaction between qubit $N$ and mediator M, and the desired
entangled state of the $(N+1)$-partite qubit M+qubit
1+$\cdots$+qubit $N$ system is generated, provided that the
interaction times, $\tau_j$ (with $j=1,2,\ldots,N$), and/or the
coupling constants, $g_j$, are accurately selected.
\end{itemize}

\begin{figure}
\includegraphics[width=0.5\textwidth]{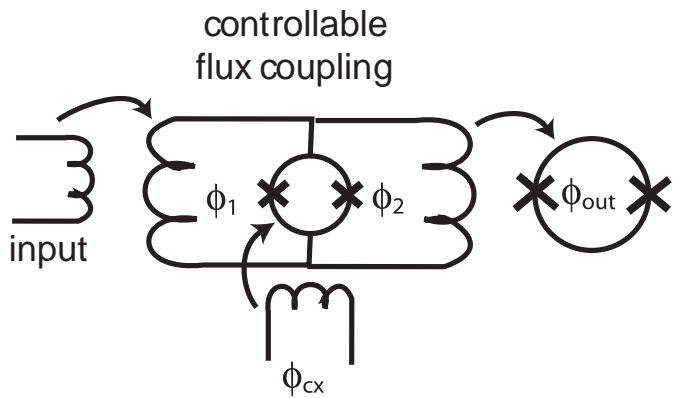}
\caption{Sketch of a controllable flux coupling (CFC) circuit. A
signal flux is applied on the left side of the CFC and its
response is read out/coupled to the SQUID on the right. The
control of the transmitted flux $\phi_\text{out}$ is achieved by
modulating the flux
$\phi_{cx}$.\cite{cosmellitransf}}\label{transfo}
\end{figure}

\begin{figure}
\includegraphics[width=0.5\textwidth]{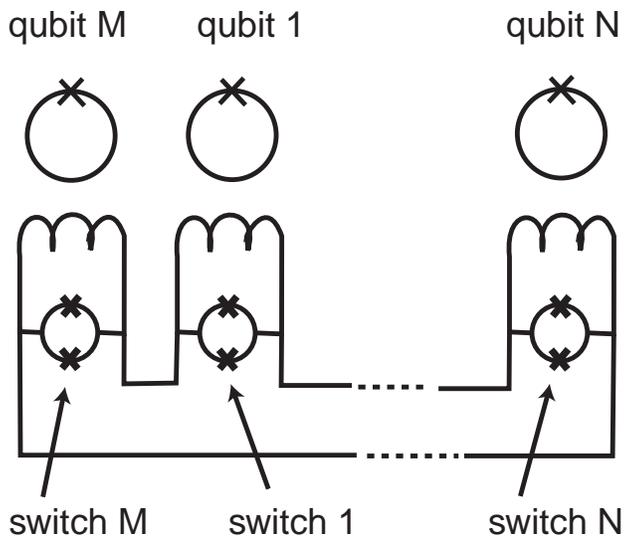}
\caption{Sketch of the scheme for the controlled coupling of more
flux qubits proposed by Cosmelli
\textit{et~al.}\cite{cosmellitransf}} \label{chain}
\end{figure}

At this stage, it is important to consider a realistic
experimental setup by which it is possible to implement our
scheme. We begin by considering an inductive mediator-qubit
coupling realized by means of a superconducting transformer with
variable flux transfer function as in the paper of Cosmelli
\textit{et~al.}\cite{cosmellitransf} They show indeed that, by
using a superconducting flux transformer modified with the
insertion of a small dc-SQUID, it is possible to control the flux
transfer function and therefore the inductive coupling constant,
by modulating via an externally applied magnetic flux $\phi_{cx}$
 the critical
current of the dc-SQUID (see Fig.\ \ref{transfo}). More in detail,
they prove that the transformer can operate between two states
with very different behavior: the ``off'' state where the transfer
ratio $\mathcal{R}= \phi_\text{out}/\phi_\text{in}$ is minimum
($\sim0$--$0.1$) and the ``on'' state with a transfer function
ratio which may be also larger than 1. Under such conditions, it
is possible to conceive an experimental scheme, like that depicted
in Fig.\ \ref{chain}, where the coupling between mediator M and
the $j$th qubit may be effectively turned on by adjusting the
control fluxes of the relative ``switches'' in such a way that one
obtains $g_j\neq 0$ with all the other ``switches'' kept in the
off state.

Similarly, Plourde \textit{et~al.}\ propose to adjust
the coupling
strength characterizing the interaction between two 3JJ flux
qubits by changing the critical current of a dc-SQUID (which is
coupled to each of these two qubits), finding that their coupling
constant can be changed continuously from positive to negative
values (and enables cancellation of the direct mutual inductive
coupling if the two qubits are not spatially
separated).\cite{plourde}

Adopting one of these experimental coupling setups, it is
therefore possible to realize the step by step scheme in the
system of $N+1$ qubits whose Hamiltonian during each of the
aforementioned steps assumes the following form:
\begin{equation}
H_{Mj}=H_0+H_{Mj}'=H_M+H_1+H_2+\cdots+H_N+H_{Mj}'\quad(j=1,2,\ldots,N).
\end{equation}
We observe that the structure of the multipartite Hamiltonian
during each step allows us to simplify the study of its dynamics
by confining ourselves to the analysis of the dynamics of the
bipartite mediator-SQUID-$1$ system during the time interval
$0<t<\tau_1$, to that of the mediator-SQUID-$2$ system during the
second period $\tau_1<t<\tau_1+\tau_2$, and so on (of course,
provided that the free evolution of the other qubits is carefully
taken into account).

Moreover, by assuming that all the qubits and entanglement
mediator M have a common energy gap $\omega=\omega_i$, $\forall\;
i\equiv\{M,1,2,\ldots,N\}$, due to the preponderance of
rotating-wave terms of the interaction Hamiltonians with respect
to the counter-rotating ones, it is not difficult to convince
oneself that, during each step, the system dynamics is dominated
by the bipartite Hamiltonian
\begin{equation}\label{reduced}
H_{Mj}^\text{RWA}=H_M+H_j+H_{Mj}^{\text{RWA}\prime}
=\frac{\hbar}{2}\omega_M\sigma_z^{(M)}
+\frac{\hbar}{2}\omega_j\sigma_z^{(j)}
+g_j(\sigma_+^{(M)}\sigma_-^{(j)}+\sigma_-^{(M)}\sigma_+^{(j)}),
\qquad\omega_M=\omega_j=\omega,
\end{equation}
describing the rotating-wave coupling between the mediator and the
$j$th qubit. In the following section, we will demonstrate the
validity of this assumption. It is interesting, however, to note
that Hamiltonian \eqref{reduced} can be exactly implemented if the
detuning $\Delta=\omega_M-\omega_j$ between the two flux (3JJ)
qubits to be coupled is chosen sufficiently large so that
initially each of them can be treated independently. As recently
shown by Liu \textit{et~al.},\cite{norideph} in fact, this gap can
be nullified by applying to one of the two SQUID loops a time
dependent magnetic flux
$\phi_x^{(\ell)}(t)=A_\ell\cos\omega_c^{(\ell)}t$ (with $\ell=M$
or $j$) satisfying the condition $\Delta\pm\omega_c^{(\ell)}=0$
when $\Delta\lessgtr0$. This means that, by considering the
reduced bias flux of each qubit close to $\phi_0/2$ and the
aforementioned frequency-matching condition, it is possible to
implement  the interaction Hamiltonian
$H_{Mj}^\text{int}=g_j(\sigma_+^{(M)}\sigma_-^{(j)}+\sigma_-^{(M)}\sigma_+^{(j)})$,
where $g_j$ can be controlled for instance by tuning the amplitude
$A_\ell$ of the applied time-dependent magnetic flux.

\section{W-State Generation}
\label{sec:Analysis} Now, we analyze the dynamics of the system
and demonstrate that the scheme introduced in the previous section
actually generates an entangled W state of a tripartite system. We
show further that it can be extended to the case of a larger
number of qubits.

We begin our analysis by  looking at the eigensolutions of
Hamiltonian \eqref{reduced}, which is represented in the basis
$\{\ket{01}_{Mj},\ket{10}_{Mj},\ket{00}_{Mj},\ket{11}_{Mj}\}$ by
the following simple diagonal block form:
\begin{equation}
H_{Mj}^\text{RWA}=\left(
\begin{array}{c c c c}
0   & g_j &              0 & 0\\
g_j &  0  &              0 & 0\\
0   &  0  &  -\hbar \omega & 0\\
0   &  0  &              0 & \hbar \omega
\end{array}
\right).
\end{equation}
These blocks describe three dynamically separate subspaces: the first with characteristic frequency $g_j/\hbar$ characterizing the appearance of the entanglement between the degenerate states $\ket{01}_{Mj}$ and $\ket{10}_{Mj}$, and the other ones describing the fact that the two states $\ket{00}_{Mj}$ and $\ket{11}_{Mj}$ evolve freely.
We easily find that the eigenstates of $H_{Mj}^\text{RWA}$ are
\begin{equation}
\ket{u_{1j}}=\frac{1}{\sqrt{2}}
[\ket{10}_{Mj}-\ket{01}_{Mj}],\quad
\ket{u_{2j}}=\frac{1}{\sqrt{2}}
[\ket{10}_{Mj}+\ket{01}_{Mj}],\quad
\ket{u_{3j}}=\ket{00}_{Mj},\quad
\ket{u_{4j}}=\ket{11}_{Mj},
\end{equation}
with eigenvalues
\begin{equation}
\lambda_{1j/2j}=\mp g_j,\qquad
\lambda_{3j/4j}=\mp\hbar\omega.
\end{equation}

\subsection{Generation of entangled W states of the tripartite ``M+1+2'' system}
By exploiting the knowledge of the eigensolutions of Hamiltonian $H_{Mj}^\text{RWA}$, it is possible to follow step by step the dynamics of the three-qubit system characterized by the one by one interaction of mediator M with qubits 1 and 2.
We choose as an initial condition the state
\begin{equation}\label{ini}
\ket{1_M0_10_2}.
\end{equation}
During the first step, we switch on the interaction between M and
1 and we let them interact for a time $\tau_1$, while 2 evolves
freely. This process generates the state
\begin{equation}
e^{-iH_{M1}\tau_1/\hbar}\ket{1_M0_10_2}
=e^{-i(H_0+H_{M1}^\text{RWA}{}')\tau_1/\hbar}\ket{1_M0_10_2}
=e^{i\omega\tau_1/2}[
\cos\theta_1 \ket{1_M0_10_2}-i\sin\theta_1\ket{0_M1_10_2}
],
\end{equation}
where $\theta_1\equiv g_1\tau_1/\hbar$. Next, by turning on the
interaction between M and 2 at $t=\tau_1$ and by allowing qubit 1
evolves freely during the second step (i.e.\ during the interval
of time $\tau_1<t<\tau_1+\tau_2$), we obtain:
\begin{align}
\ket{\varphi_2}
&=e^{-iH_{M2}\tau_2/\hbar}
e^{-iH_{M1}\tau_1/\hbar}\ket{1_M0_10_2}\nonumber\\
&=e^{i\omega(\tau_1+\tau_2)/2}
[\cos\theta_1\cos\theta_2\ket{1_M0_10_2}
-i\sin\theta_1\ket{0_M1_10_2}
-i\cos\theta_1\sin\theta_2\ket{0_M0_11_2}].
\label{2step}
\end{align}
Equation \eqref{2step} clearly shows that, by adjusting $\theta_j\equiv g_j \tau_j/\hbar$ ($j=1,2$) so that
\begin{equation}\label{cond}
|\sin\theta_1|=\frac{1}{\sqrt{3}}\quad\text{and}\quad
|\cos\theta_2|=\frac{1}{\sqrt{2}},
\end{equation}
a tripartite W state is generated.
If we choose $\theta_2=\pi/4$ and $\theta_1$ such that $\sin\theta_1=1/\sqrt{3}$ and $\cos\theta_1=\sqrt{2/3}$, for instance, we get
\begin{equation}\label{w}
\ket{W_2}
=\frac{1}{\sqrt{3}}e^{i\omega(\tau_1+\tau_2)/2}
[\ket{1_M0_10_2}-i\ket{0_M1_10_2}-i\ket{0_M0_11_2}].
\end{equation}
It is worth noting that we obtain this state by adjusting the coupling energies $g_j$ during the aforementioned steps and/or by tuning the interaction times $\tau_j$.

\subsection{Generation of entangled W states of the multipartite ``M+1+2+$\cdots$+$N$'' system}
Within this framework, it is possible to look at the possibility of applying the same techniques in order to generate an entangled state of a multipartite ``M+1+2+$\cdots$+$N$'' system with $N>2$.
As described in Sec.\ III, we consider an entanglement mediator M in interaction one by one with $N$ qubits.
If the system is prepared at $t=0$ in the factorized state
\begin{equation}
\ket{1_M0_10_2\ldots0_N},
\end{equation}
after a straightforward approach it is easy to show that at the end of the $N$th step (namely at $t=t_N\equiv \sum_{j=1}^N\tau_j$) it can be described in terms of the  state
\begin{multline}
\ket{\varphi_N}=e^{i(N-1)\omega t_N/2}
[\cos \theta_1\cos \theta_2\cdots\cos\theta_N\ket{1_M0_10_2\ldots0_N}\\
-i\sin\theta_1\ket{0_M1_10_2\ldots0_N}
-\cdots-i\cos\theta_1\cos \theta_2\cdots\cos \theta_{k-1}\sin\theta_k\ket{0_M0_10_2\ldots1_k\ldots0_N}\\
-\cdots-i\cos\theta_1\cos \theta_2\cdots\cos\theta_{N-1}\sin\theta_N\ket{0_M0_10_2\ldots1_N}
].
\end{multline}
Assuming that it is possible to control the interaction time
$\tau_j$ of each qubit with mediator M and/or their coupling
constants $g_j$ so that
\begin{eqnarray}
\sin\theta_j=\frac{1}{\sqrt{N-j+2}}\quad\text{and}\quad
\cos\theta_j=\sqrt{\frac{N-j+1}{N-j+2}},
\label{eqn:TuningN}
\end{eqnarray}
we finally find that the generalized W entangled state
\begin{equation}
\ket{W_N}
=\frac{e^{i(N-1)\omega t_N/2}}{\sqrt{N+1}}
[\ket{1_M0_10_2\ldots0_N}
-i\ket{0_M1_10_2\ldots0_N}
-i\ket{0_M0_11_2\ldots0_N}
\cdots-i\ket{0_M0_10_2\ldots1_N}
]
\label{wn}
\end{equation}
of the $(N+1)$-partite system is created.

\subsection{Estimation of the effect of counter-rotating terms}
At this stage, we wish to test the validity of the rotating-wave
approximation (RWA) performed at the end of Sec.\ \ref{sec:Scheme}.
To this end, we analyze the fidelity of the system state
$\ket{\psi_N}$ calculated without performing the RWA on the
Hamiltonian model with respect to the target state \eqref{wn}, i.e.\
$\mathcal{F}_N=|\braket{W_N}{\psi_N}|$.

Let us first look at the fidelity for the tripartite W state.
After a straightforward calculation, we find that, by following
the aforementioned two-step procedure, the
$\sigma_x^{(M)}\sigma_x^{(j)}$ couplings generate the state
\begin{multline}
\ket{\psi_2}
=\frac{1}{\sqrt{3}}e^{i\omega(\tau_1+\tau_2)/2}
\left\{
\ket{1_M0_10_2}-i\ket{0_M0_11_2}-ie^{-i\omega\tau_2}\left[
\left(
\cos\chi_2
+i\frac{\hbar\omega}{\sqrt{g_2^2+\hbar^2 \omega^2}}\sin\chi_2
\right)\ket{0_M1_10_2}
\right.\right.\\
\left.\left.
-i\frac{g_2}{\sqrt{g_2^2+\hbar^2\omega^2}}\sin\chi_2\ket{1_M1_11_2}
\right]
\right\}
\label{eqn:Wxx}
\end{multline}
with the same $\theta_1$ as that for \eqref{w}, where
$\chi_2=\sqrt{g_2^2/\hbar^2+\omega^2}\tau_2$. Its fidelity
$\mathcal{F}_2=0.999957$, calculated with
$g_2/\hbar\approx0.5\,\text{GHz}$, $\omega\approx10\,\text{GHz}$
(in agreement with the currently available experimental values),
confirms that during each step the system dynamics is dominated by
the bipartite Hamiltonian \eqref{reduced} describing the
rotating-wave coupling between the mediator and the $j$th qubit.

For the ($N+1$)-partite system, the generated state reads
\begin{multline}
\ket{\psi_N}
=\frac{e^{i(N-1)\omega t_N/2}}{\sqrt{N+1}}\,\Biggl[
\ket{1_M0_10_2\ldots0_N}-i\ket{0_M0_10_2\ldots1_N}\\
-i\sum_{k=1}^{N-1}\left\{
\prod_{j=k+1}^Ne^{-i\omega\tau_j}\left(
\cos\chi_j
+i\frac{\hbar\omega}{\sqrt{g_j^2+\hbar^2\omega^2}}\sin\chi_j
\right)
\right\}\ket{0_M0_10_2\ldots1_k\ldots0_N}\\
+\text{(states orthogonal to $\ket{W_N}$)}
\Biggr]
\label{eqn:WNxx}
\end{multline}
with the tuning (\ref{eqn:TuningN}), where
$\chi_j=\sqrt{g_j^2/\hbar^2+\omega^2}\tau_j$, and the fidelity is
given by
\begin{equation}
\mathcal{F}_N
=\frac{1}{N+1}\left|
2+\sum_{k=1}^{N-1}
\prod_{j=k+1}^Ne^{-i\omega\tau_j}\left(
\cos\chi_j
+i\frac{\hbar\omega}{\sqrt{g_j^2+\hbar^2\omega^2}}\sin\chi_j
\right)
\right|,
\label{eqn:FNxx}
\end{equation}
which results in
\begin{equation}
\mathcal{F}_3=0.999873,\quad
\mathcal{F}_4=0.999601,\quad\ldots,\quad
\mathcal{F}_{10}=0.997561,\quad\ldots,\quad
\mathcal{F}_{20}=0.993959,\quad\text{etc.}
\end{equation}
when $g_1/\hbar=\cdots=g_N/\hbar\approx0.5\,\text{GHz}$ and
$\omega\approx10\,\text{GHz}$. The physical meaning of such a list
of values is that we might implement our scheme using up to $20$
qubits maintaining a very good level of fidelity of the state
given by Eq.\ \eqref{eqn:WNxx} with respect to the target state
expressed by Eq.\ \eqref{wn}.

\section{Conclusions}
\label{sec:Conclusion} In this paper, we have discussed a scheme
for the generation of entangled W states among three Josephson
(eventually spatially separated) flux qubits as well as its
generalization to the case of $N+1$ qubits. The success of the
scheme relies on the possibility of realizing controllable
couplings between the qubits and entanglement mediator M and on
the possibility of preparing their initial quantum state. It
should be stressed that no conditional measurement are required
but we have to tune the coupling energy and/or the interaction
time between each qubit and the mediator.

Final considerations are devoted to the analysis of some aspects
of decoherence problem in our system. In particular, we wish to
verify if the operation duration (e.g.\ the time necessary to
perform the desired quantum process) is small enough with respect
to the decoherence time. We find, on the one hand, that the
eigenfrequency of a Josephson qubit $\omega$ is of the order of
$10\,\text{GHz}$ and that correspondingly
$g_j/\hbar\approx0.5\,\text{GHz}$. Under such conditions, the
length of each step (during which only a fraction of a Rabi
oscillation takes place) is at most of the order of
$\hbar/g_j\approx 2\,\text{ns}$ and consequently the whole process
in the case, for instance, of a tripartite system lasts
approximatively $4\,\text{ns}$. On the other hand the relaxation
and decoherence times, $T_1$ and $T_2$, of a superconducting flux
qubit have been coarsely estimated and are in the range
$1$--$10\,\mu\text{s}$.\cite{noritoday,cosmel} Therefore this
means that a number of operations can be performed before
decoherence occurs and that, in principle, our quantum information
processing can be exploited for the experimental generation of
entangled W states among more than three qubits.

\begin{acknowledgments}
This work is partly supported by the bilateral Italian--Japanese
Projects II04C1AF4E, tipology C, on ``Quantum Information,
Computation and Communication'' of the University of Bari (D.M.
05/08/2004, n. 262, Programmazione del sistema universitario
2004/2006, Art. 23- Internationalizzazione) and 15C1 on ``Quantum
Information and Computation'' of the Italian Ministry for Foreign
Affairs, by the Grant for The 21st Century COE Program ``Holistic
Research and Education Center for Physics of Self-Organization
Systems'' at Waseda University and the Grant-in-Aid for Scientific
Research on Priority Areas ``Dynamics of Strings and Fields''
(No.\ 13135221) both from the Ministry of Education, Culture,
Sports, Science and Technology, Japan, and by a Grant-in-Aid for
Scientific Research (C) (No.\ 14540280) from the Japan Society for
the Promotion of Science.

\end{acknowledgments}


\end{document}